%% file: main.tex
\documentclass[final, 1p, times, authoryear]{elsarticle}

\usepackage{amssymb}
\usepackage{amsthm}
\usepackage[colorlinks=true, allcolors=blue]{hyperref}     
\usepackage{bm}
\usepackage[ruled,vlined]{algorithm2e}
\usepackage{algorithmic}
\usepackage{amsmath}
\usepackage{array}
\usepackage{multirow}
\usepackage{xcolor}   
\usepackage[subpreambles=true]{standalone}

\begin{document}



\begin{frontmatter}
\title{Application of the Modular Bayesian Approach for Inverse Uncertainty Quantification in Nuclear Thermal-Hydraulics Systems}



\author{Chen Wang\corref{cor1}\fnref{label1}}
\cortext[cor1]{Corresponding author. Email address: chenw3@illinois.edu}

\affiliation[label1]{organization={Department of Nuclear, Plasma and Radiological Engineering, University of Illinois Urbana Champaign}}

\begin{abstract}

In the framework of BEPU (Best Estimate plus Uncertainty) methodology, the uncertainties involved in the simulations must be quantified to prove that the investigated design is acceptable. The output uncertainties are usually calculated by propagating input uncertainties through the simulation model, which requires knowledge of the model input uncertainties. However, in some best-estimate Thermal-Hydraulics (TH) codes such as TRACE, the physical model parameters used in empirical correlations may have large uncertainties, which are unknown to the code users. Therefore, obtaining uncertainty distributions of those parameters becomes crucial if we want to study the predictive uncertainty or output sensitivity. 

In this study, we present a Modular Bayesian approach that considers the presence model discrepancy during Bayesian calibration. Several TRACE physical model parameters are selected as calibration parameters in this work. Model discrepancy, also referred to as model inadequacy or model bias, accounts for the inaccuracy in computer simulation caused by underlying missing/insufficient physics, numerical approximation errors, and other errors of a computer code, even if all its parameters are fixed at their ``true'' values. Model discrepancy always exists in computer models because they are reduced representations of the reality. The consideration of model discrepancy is important because it can help avoid the “overfitting” problem in Bayesian calibration. This paper uses a set of steady-state experimental data from PSBT benchmark and it mainly aims at: (1) quantifying the uncertainties of TRACE physical model parameters based on experiment data; (2) quantifying the uncertainties in TRACE outputs based on inversely quantified physical model parameters uncertainties. 

\end{abstract}

\end{frontmatter}


\input{1intro}
\input{2sec}
\input{3sec}
\input{4sec}

\input{5sec}
\input{6sec}


\bibliographystyle{elsarticle-harv} 
\bibliography{main}



\end{document}

%% file: 1intro.tex
\section{Introduction}
\label{introduction}

In recent advancements, the domain of Nuclear Thermal Hydraulics (TH) has increasingly embraced computational simulations as a cornerstone for predicting the behavior of nuclear reactor systems. These simulations are imperative for maintaining the safety and operational efficiency of reactors, yet they are marred by inherent uncertainties due to the complex nature of the physical phenomena they attempt to model. The introduction of time-dependent aspects further compounds these uncertainties, adding layers of complexity to the simulations. To navigate these challenges, Inverse Uncertainty Quantification (IUQ) has been identified as a critical methodology, focusing on quantifying the uncertainties tied to Physical Model Parameters (PMPs) within these simulations.

The trajectory of IUQ has been significantly influenced by the evolving landscape of machine learning (ML) and artificial intelligence (AI), which have enhanced the accuracy and reliability of simulation models across diverse sectors. These technologies have proven their effectiveness by addressing complex challenges in various fields, for example, healthcare (\cite{dong2021influenza,dong2021semi,chen2019claims}), agriculture (\cite{wu2022optimizing, wu2024new}), transportation (\cite{ma2022application,meng2022comparative,li2023exploring}), clinical research (\cite{xue2021use,xue2022perioperative}), signal processing (\cite{hu2022dan,li2023nst, liu2021measuring}), structural health monitoring (\cite{liu2023intelligent}), reliability engineering (\cite{chen2020optimal, wang2024optimal, chen2017multi, chen2020some}), industrial engineering (\cite{chen2018data, li2023applying, chen2018data, chen2023recontab, wu2024switchtab}), and artificial intelligence (\cite{liu2019dapred, liu2023stationary, liu2024cliqueparcel, lai2024language, lai2024adaptive, zhi2017claimverif}). The successful application of ML/AI in these domains underscores their potential and offers valuable insights for propelling IUQ forward in the nuclear industry.

Traditional IUQ approaches, predominantly using single-level Bayesian models, have provided significant insights into steady-state TH systems. However, they exhibit limitations when applied to applications with various experimental conditions and large datasets. Key challenges include handling the high variability of PMPs under dynamic experimental conditions and avoiding over-fitting due to unknown model discrepancies or outliers. Recent advancements in hierarchical Bayesian models ~\cite{wang2023inverse,wang2023scalable} have shown promise in addressing these issues in nuclear TH systems.

This study aims to fill this gap by introducing a Modular Bayesian model tailored for IUQ in nuclear TH systems where mdoel bias is significant. The Modular Bayesian approach \cite{wang2018ans, wu2017inverse2, wu2017inverse} considers the presence model discrepancy during Bayesian calibration. Several TRACE physical model parameters are selected as calibration parameters in this work. Model discrepancy, also referred to as model inadequacy or model bias, accounts for the inaccuracy in computer simulation caused by underlying missing/insufficient physics, numerical approximation errors, and other errors of a computer code, even if all its parameters are fixed at their “true” values. Model discrepancy always exists in computer models because they are reduced representations of the reality. The consideration of model discrepancy is important because it can help avoid the “overfitting” problem in Bayesian calibration \cite{wangnureth1, wangnureth3}. 

This paper uses a set of steady-state experimental data from PSBT benchmark, and it mainly aims at: (1) quantifying the uncertainties of TRACE physical model parameters based on experiment data; (2) quantifying the uncertainties in TRACE outputs based on inversely quantified physical model parameters uncertainties. 

The rest of the paper is organized as follows. Section \ref{sec2} will give an overview of Bayesian model for IUQ. Section \ref{sec3} will introduce the overview of the PSBT benchmark and TRACE modeling. Section \ref{sec4} will introduce the surrogate models for time-dependent problems, and then in Section \ref{sec5}, the IUQ framework is applied to a case study for TRACE physical model parameters using the PSBT benchmark data. Section \ref{sec6} will be the summary.

%% file: 2sec.tex
\section{Bayesian Framework for Inverse Uncertainty Quantification}
\label{sec2}

A key assumption in most of the Bayesian-based Inverse UQ framework is the model updating equation. Following the work of Kennedy and O’Hagan ~\cite{kennedy2001bayesian}, we represent the relationship between the computer model outputs $\bm y^M(\bm x, \bm \theta)$ and the observations $\bm y^E(\bm x)$ in the equation:

\begin{equation}
\label{eqa:21}
    \bm y^E(\bm x) = \bm y^M(\bm x, \bm \theta) + \bm \delta(\bm x) + \bm \epsilon 
\end{equation}

where $\bm \epsilon $ is a vector of observation error, and we assume $\bm \epsilon$ is independent and identically distributed as $\mathcal{N}(0,\sigma_{exp}^2)$. It should be noted that this assumption may not always hold in reality, more details about this assumption will be discussed in Chapter \ref{chap6}. 

$\bm \delta(\bm x)$ is the model discrepancy term, which is caused by incomplete or inaccurate physics employed in the model. The discrepancy term $\bm \delta(\bm x)$ is only a function of the control parameters $\bm x$, which is a consequence of the fundamental difference between $\bm x$ and $\bm \theta$. Here, the parametric uncertainty is derived from the $\bm \theta$ parameter, and other forms of uncertainties are incorporated in the model discrepancy term. 

Following the model updating equation, the posterior PDF of the calibration parameter can be found using Bayes' rule:

\begin{equation}
\label{eqa:22}
    p(\bm \theta | \bm y^E, \bm y^M) \propto p(\bm y^E, \bm y^M |\bm \theta) \cdot p(\bm \theta)
\end{equation}

where $p(\bm \theta)$ is the prior distribution of the calibration parameter, and $p(\bm y^E, \bm y^M |\bm \theta)$ is the likelihood function. From equation \ref{eqa:21}, we know that $\bm \epsilon = \bm y^E(\bm x) - \bm y^M(\bm x, \bm \theta) - \bm \delta(\bm x) $ follows a multivariate normal distribution. So the posterior can be written as:

\begin{equation}
    \label{eqa:23}
    p(\bm \theta | \bm y^E, \bm y^M) \propto \frac{1}{\sqrt{|\bm \Sigma_t|}} \exp \bigg[-\frac{1}{2}[\bm y^E-\bm y^M - \bm \delta]^T \bm{\Sigma}_t^{-1}[\bm y^E-\bm y^M - \bm \delta] \bigg] \cdot p(\bm \theta)
\end{equation}

where the covariance matrix $\bm{\Sigma}_t$ is defined as: 
\begin{equation}
    \label{eqa:sigma_total}
    \bm{\Sigma}_t = \bm \Sigma_{exp} + \bm \Sigma_{\delta} + \bm \Sigma_{code}
\end{equation}
where $\bm \Sigma_{exp}$ is the experimental measurement uncertainty, $\bm \Sigma_{\delta}$ is the model uncertainty due to inaccurate underlying physics, and $\bm \Sigma_{code}$ is the model uncertainty introduced by surrogate models when surrogate model is used as an approximation of the original TH code. It should be noted that the treatment of $\bm \Sigma_t$ here is a traditional method and has been widely used in previous work (\cite{wu2017inverse} \cite{wu2018kriging} \cite{wangnureth1} \cite{wang2019gaussian}).

%% file: 3sec.tex
\section{Validation of TRACE based on Steady-State Void Fraction Measurements in PSBT Benchmark}
\label{sec3}

\subsection{PSBT Benchmark}

OECD/NRC benchmark based on NUPEC PWR subchannel and bundle tests (PSBT) is designed for validation purposesof void distribution in subchannel and PWR bundle and pre-diction of departure form nucleate boiling Uncertainty Quantification for Steady-State PSBT Benchmark using Surrogate Models. 
The void distribution benchmark in PSBT includes transient bundle benchmark, which can be applied to system TH codes to assess their capabilities of predicting the void generation during transients (\cite{specifications2010oecd}). The experimental data in these transients include X-ray densitometer measurements of void fraction (chordal averaged) at three axial elevations. The averaging is over the four central subchannels. Data is collected for four transient scenarios: Power Increase (PI), Flow Reduction (FR), depressurization (DP), Temperature Increase (TI), and at three different assembly types 5, 6, and 7. All 5, 6, and 7 assemblies are $5 \times 5$ rob bundles while 5 and 6 have typical cells and 7 has thimble cells. They also have different axial and radial power distributions. The PSBT benchmark has been used extensively in many IUQ related applications (\cite{borowiec2017uncertainty,wang2018ans})

\begin{figure}[!h]
    \centering
    \includegraphics[width = 0.8\textwidth]{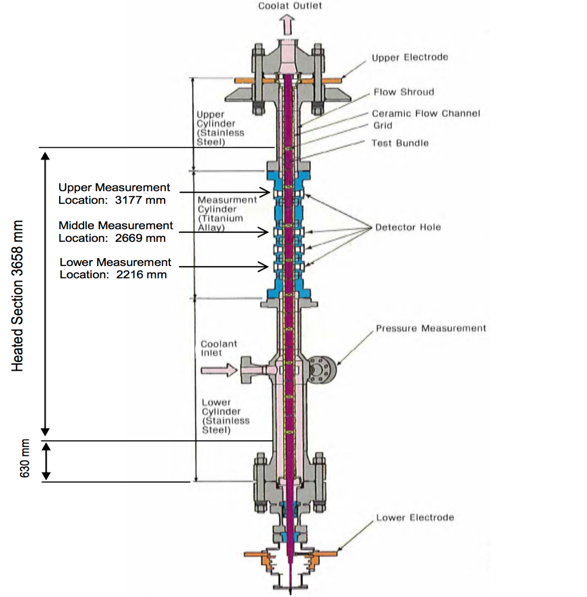}
    \caption{Test Section for PSBT Rod Bundle Void Distribution Measurement~\cite{specifications2010oecd} }
    \label{fig:psbt_test}
\end{figure}

In this work, we will use the steady-state bundle void distribution measurements in the benchmark. The corresponding test section is shown in Figure \ref{fig:psbt_test}. As we can see in the figure, an electrically heated rod bundle is used to simulate a partial section and full length of a PWR fuel assembly, and the coolant flows from the bottom of the pressure vessel up through the test assembly. The experiment was conducted at different assembly types and different boundary conditions. The average void fractions data are measured at three different locations at 3177 mm, 2699 mm, 2216 mm, within the effective heated length of 3658 mm. The void fraction of the gas-liquid two-phase flow was converted from the density of the flow, which was measured by a gamma-ray transmission method. The measurement noise (uncertainty) of the steady-state void fraction data was reported to be $4\%$ void. 

\subsection{TRACE Simulation and Physical Model Parameters}

TRACE is a best-estimate reactor system code developed by U.S. Nuclear Regulatory Commission for analyzing both transient and steady-state neutronics-thermal-hydraulic behavior in light water reactors~\cite{bajorek2008trace}. The hydraulic module of TRACE is based on a two-fluid six-equation model, solving the conservation equations of mass, momentum, and energy for the liquid and vapor phases in the coolant.

Major challenges for current system TH modeling are caused by our lack of understanding and proper techniques to model the interaction mechanism at the interface between the liquid and vapor phases. Empirical correlations are widely used to model the interfacial transfer mechanism (especially the interfacial momentum transfer). Consequently, substantial uncertainties can be propagated from these correlations to predictions of the two-phase tow-fluid model\cite{wu2017inverse}.

\begin{figure}[!h]
    \centering
    \includegraphics[width = 0.8\textwidth]{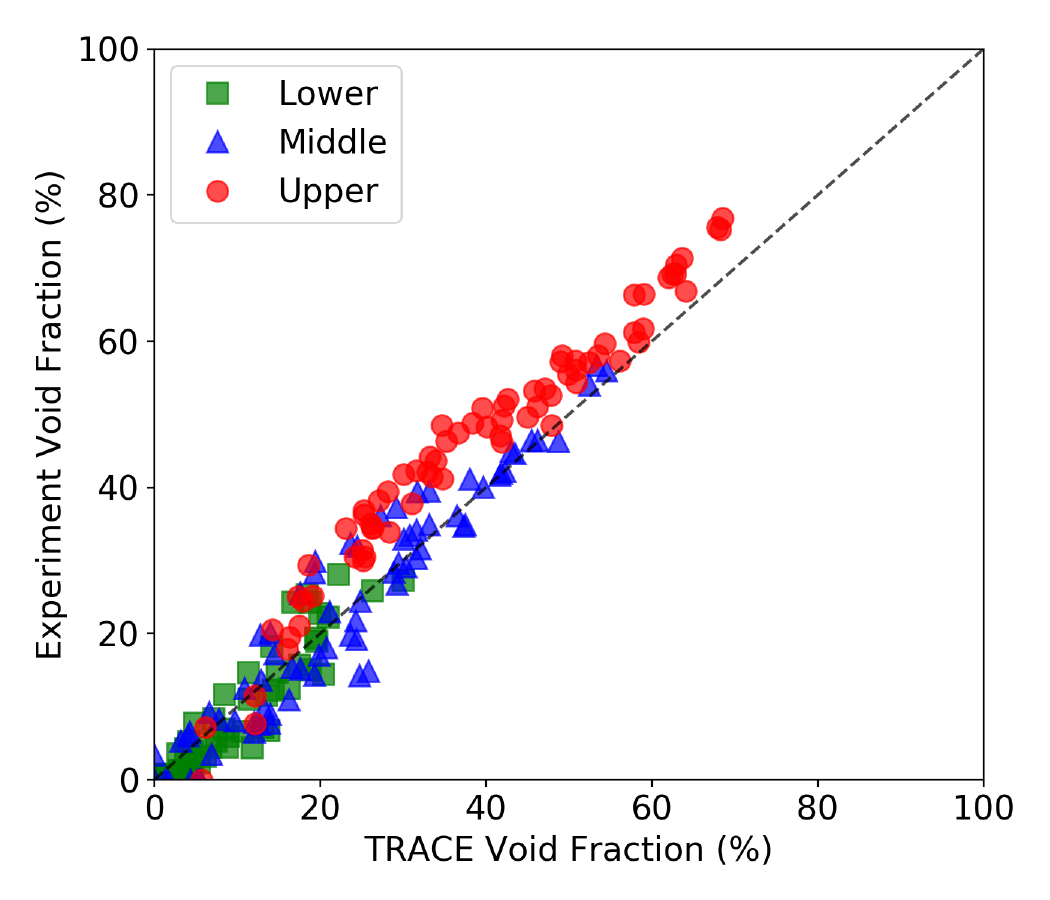}
    \caption{Comparison of the predicted void fractions by TRACE and experimental measurements. PSBT benchmark test assembly 5}
    \label{fig:compare_psbt_1}
\end{figure}

A TRACE model is built according to the test assembly geometry. All 74 cases in PSBT bundle test series 5 are selected in this study. These 74 test cases have the same assembly type and geometry, but different boundary conditions (pressure, coolant inlet temperature, mass flow rate and power). A comparison between the simulation results of TRACE (using nominal values of all the uncertain physical model parameters) and the experimental data is shown in Figure~\ref{fig:compare_psbt_1}.

Void fractions at `Upper', `Middle' and `Lower' measurement locations are indicated by three different colors/shapes in Figure~\ref{fig:compare_psbt_1}. We can see that the simulation results are generally consistent with experimental measurements. However, it is obvious that TRACE tends to under-predict the void fraction at the Upper measurement location, which indicates the potential existence of model inadequacy. 

For the conservation equations in the TRACE two-phase flow model, closure laws or constitutive relations are required to obtain a closed solution, where some parameters such as interfacial drag coefficient and liquid and vapor wall drag coefficient need to be modeled. These parameters along with some other physical model parameters can be adjusted by a multiplicative factor in TRACE, allowing users to conduct sensitivity and uncertainty analysis for those parameters. This paper will treat these physical model parameters (more specifically, their multiplication factors) as uncertain inputs and inversely quantify their probabilistic distributions.

\section{Sensitivity Analysis}
\label{sec:sa}
Following the flowchart outlined in Figure \ref{fig:1}, SA will now be conducted to select input parameters for calibration. TRACE provides the option to adjust 36 physical model parameters by a multiplicative factor, however, not all of those parameters will be active in the PSBT bundle assembly model because some parameters involve phenomena that do not occur in PSBT benchmark, e.g. reflood. The details of these 36 physical model parameters can be found in the TRACE manual~\cite{bajorek2008trace} or in the Appendix. The aim of this part of work is to remove all the non-influential ones to reduce the unnecessary computational burden in the following emulator construction and MCMC sampling processes. 

Most of the parameters are multiplicative factors and some are additive factors, so their nominal values are $1.0$. A simple perturbation method is used to perturb each parameter in the range of $(0,5)$ while fixing other parameters. 50 uniform samples in that range are used to test the effect of this parameter on the simulated void fraction data. The resulting output variance is calculated for each parameter. The results show that most of the variances are 0 or very close to 0. Finally, eight parameters with variances larger than $10^{-3}$ are selected and shown in Table \ref{tab:parameter_describ}.

\begin{table}[h!]
    \centering
    \caption{List of 8 selected physical model parameters in TRACE}
    \begin{tabular}[c]{m{5em}  c}
    \hline
    \textbf{Parameter Number}     &  \textbf{Definition} \\
    \hline
    P1000 & Liquid to interface bubbly-slug heat transfer\\
    P1002 & Liquid to interface transition heat transfer coefficient \\
    P1008     & Single phase liquid to wall heat transfer coefficient \\
    P1012     & Subcooled boiling heat transfer coefficient   \\
    P1022     & Wall drag coefficient \\
    P1028     & Interfacial drag (bubbly/slug Rod Bundle-Bestion) coefficient\\
    P1029 & Interfacial drag (bubbly/slug Vessel) coefficient\\
    P1030 & Interfacial drag (annular/mist Vessel) coefficient\\
    \hline
    \end{tabular}
    \label{tab:parameter_describ}
\end{table}

Next, we will conduct a more accurate SA for the selected 8 parameters. Sobol' indices method is used here, and the first (main) and total Sobol' indices are shown in Figure \ref{fig:sa1}. The colors represent the corresponding indices for a certain measurement location. We can see that four out of eight have more significant influences on the model outputs, so the four parameters P1008, P1012, P1022, and P1028 are selected in the sensitivity analysis step and will be treated as uncertain inputs. It should be noted that the Sobol method in this figure is obtained by samples from only one experiment case. It is not wise nor necessary to repeat the computation for all cases because they all have the same geometry and governing physics. Several calculations of Sobol indices on randomly selected cases are completed for a sanity check, and the results show the value of the Sobol indices for the middle four parameters may be different, but the other four parameters always show negligible sensitivities. So it is safe to only select 'P1008', 'P1012', 'P1022', 'P1028' as uncertain parameters for the following calibration process. 

\begin{figure}[!h]
    \centering
    \includegraphics[width = \textwidth]{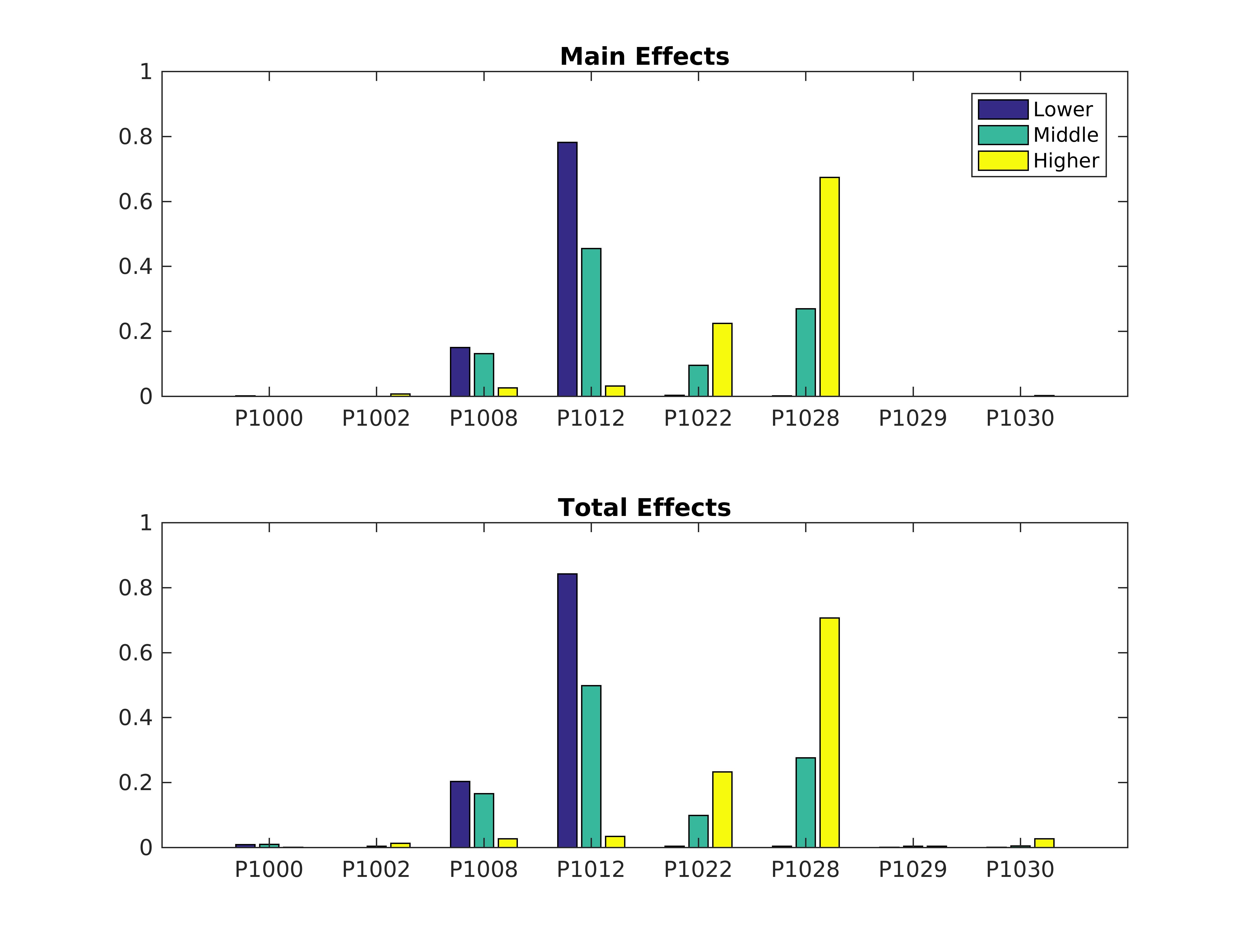}
    \caption{Sobol indices for the 8 parameters on 3 VF outputs. PSBT test assembly 5, case 1. }
    \label{fig:sa1}
\end{figure}

Including non-sensitive parameters in  Bayesian calibration is generally not a dangerous thing, because we can expect the resulting posterior to be very similar to the prior, which is wide and non-informative, but it would not affect other parameters. During the MCMC sampling stage, including non-important parameters also would not significantly decrease the efficiency of the algorithm, because the parameter barely affects the acceptance rate of the proposed new points. The pain point of having many uncertain inputs is the input dimension of the surrogate model we need to build for TRACE. The number of simulations required to construct a ``equally good'' surrogate model increases exponentially with the dimensions of the computer model, which is usually referred to as ``curse of dimensionality''. So reducing the input dimension in the surrogate-based Bayesian calibration approach can significantly reduce the total number of runs of the original computer model. 

Finally, it is interesting to see how the void fraction will change as these four physical model parameters change. Figure \ref{fig:sa1} shows the change in void fraction as each physical model parameter is perturbed in the range of 0 to 12. Each line in the figure represents the void fraction value as a function of only one physical model parameter. We can see that the most significant changes occur in the range of 0 to 2, which is reasonable because the physical model parameters are multiplicative factors. We need to pay close attention the wall drag coefficient because its trend keeps decreasing in an almost constant rate, indicating that a short-range may not be enough to reflect its impacts. This phenomenon is also consistent with the definition of the wall drag coefficient.

\begin{figure}[!htbp]
    \centering
    \includegraphics[width = 0.9 \textwidth]{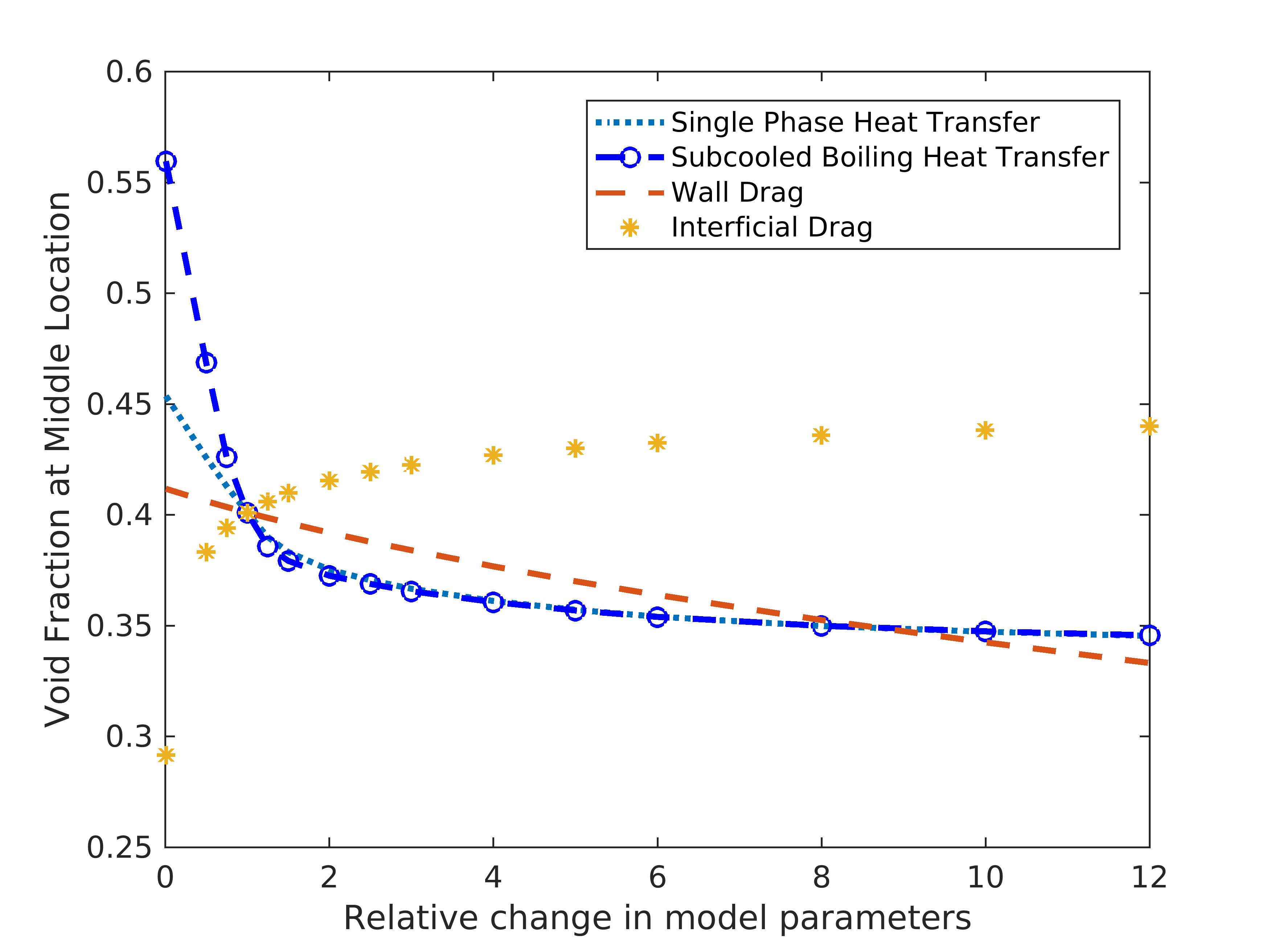}
    \caption{Effects of the selected four physical model parameters on the predicted void fraction. PSBT test assembly 5 case 1}
    \label{fig:sa1}
\end{figure}

Since the transient and the steady-state experiments share same assemblies and similar boundary conditions, it is reasonable to use the same physical model parameters as we have selected in previous similar studies (\cite{wang2017sensitivity, wang2019gaussian, wangsurrogate, wangnureth3, mc17-1, mc17-2}). The four parameters are listed in Table \ref{tab:6parameter_2}.

\begin{table}[!htbp]
    \centering
    \caption{List of 4 selected physical model parameters in TRACE}
    \begin{tabular}[c]{m{5em}  c}
    \hline
    \textbf{Parameter Number}     &  \textbf{Definition} \\
    \hline
    P1008     & Single phase liquid to wall heat transfer coefficient \\
    P1012     & Subcooled boiling heat transfer coefficient   \\
    P1022     & Wall drag coefficient \\
    P1028     & Interfacial drag (bubbly/slug Rod Bundle-Bestion) coefficient\\
    \hline
    \end{tabular}
    \label{tab:6parameter_2}
\end{table}

%% file: 4sec.tex
\section{Surrogate Model for Time-Dependent Thermal-Hydraulics Systems}
\label{sec4}

Since an obvious model discrepancy can be observed in the void fraction measurements in the upper location in Figure~\ref{fig:compare_psbt_1}, we will use the modular Bayesian approach to quantify the posterior distributions of the physical model parameters, with model discrepancy taken into account. GP for computer code (TRACE) $GP_{CC}$ and GP for model discrepancy $GP_{MD}$ will be built, respectively. Boundary conditions are considered as the control parameters $\bm x$, and the physical model parameters are considered as calibration parameters $\bm \theta$. 

\begin{figure}[!h]
    \centering
    \includegraphics[width = \textwidth]{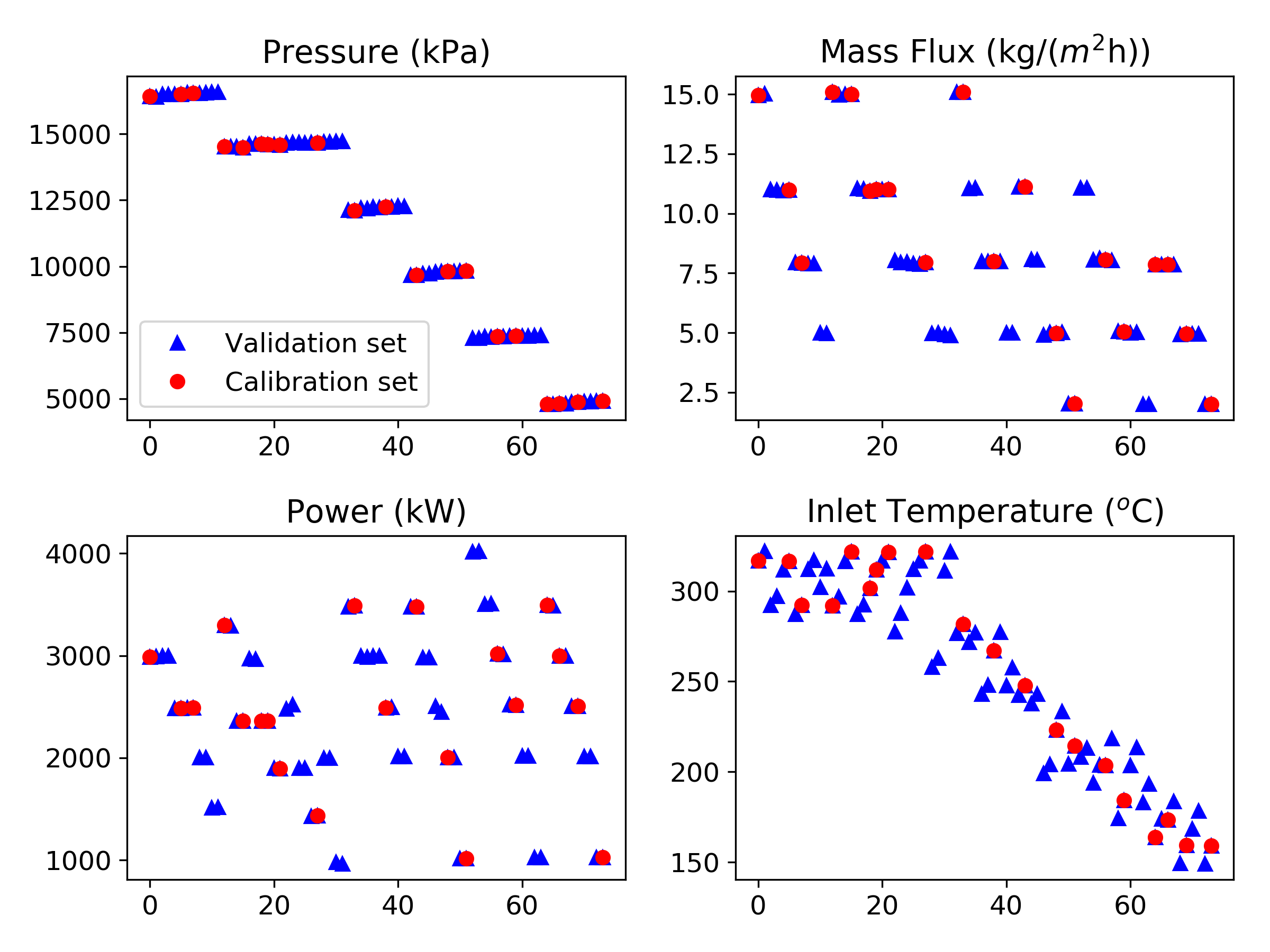}
    \caption{Boundary conditions in PSBT void distribution test assembly 5}
    \label{fig:bc}
\end{figure}

As we have explained in section \ref{sec:fmba}, the $GP_{CC}$ is built based on the simulation date $\bm y^M$ at the given $N$ points $[(\bm x_1,\bm \theta_1),(\bm x_2,\bm \theta_2),...,(\bm x_N,\bm \theta_N)]$, and the $GP_{MD}$ is built based on $[\bm y^E(\bm x_1) - \bm y^M(\bm x_1), \bm y^E(\bm x_2) - \bm y^M(\bm x_2),...,\bm y^E(\bm x_M) - \bm y^M(\bm x_M)]$. Now the question is how do we get the corresponding training samples for these two models. In a model where it is okay to adjust both the control parameter and the calibration parameter, the task would be easy because we can use the LHS method to draw random samples from the given ranges of ($\bm x, \bm \theta$) and ($\bm x$) respectively. However, in nuclear Thermal-Hydraulics, the accuracy of computer simulations cannot be guaranteed in untried points, especially when it comes to extrapolation. So it is safer to directly use existing boundary conditions as samples to train $GP_{CC}$. Luckily, various boundary conditions in the 74 cases of the PSBT test assembly 5 give us enough samples for this task. The boundary conditions in these 74 cases are shown in Figure \ref{fig:bc}.

The x-axis in Figure~\ref{fig:bc} is the case number, and the y-axes are the corresponding boundary conditions. The cases are split into two sets: validation set and calibration set. The calibration set is used to train $GP_{CC}$, and the validation set is used to train $GP_{MD}$. The reason for this treatment is that the datasets of the control variable to train the GP model for computer code and for model discrepancy must be different, otherwise, the computer code would play no role in the model updating equation because the GP model interpolates exactly. If the same dataset is used, only the measurement error term is calibrated, which is meaningless. We selected 20 cases out of 74 as the calibration set to train the GP model for computer code. The rule for the selection is that the calibration domain should be encompassed by the validation domain, otherwise extrapolation might occur and make the GP model for model discrepancy inaccurate. More cases are allocated to the validation set because more samples would guarantee a more accurate GP for the model discrepancy. Once the accuracy of the GP model can be guaranteed, more cases can be allocated to the calibration set to support our understanding about the calibration parameters. Test source allocation (TSA) is a method developed by \cite{wu2018inverse} and \cite{wu2018inverse22} for data partition tasks in this situation. It can also be done by a data partition method proposed by Morrison et al, where all possible partitions are considered to find the optimal partition (\cite{morrison2013data}).

Now $GP_{MD}$ can be trained based on the validation set, where the input is $\bm x_{val}$ and the output is $\bm y^E(\bm x_{val}) - \bm y^M(\bm x_{val})$. Note that the simulation output of $y^M(\bm x_{val})$ is run at the nominal value of $\bm \theta$, which is $1.0$. The accuracy of the c can be quantified by cross validation or leave-one-out-error. Now that $GP_{MD}$ is constructed, if we look at the predicted value of $GP_{MD} (\bm x_{cal})$, which just is the predicted model discrepancy at the calibration set, we can expect two outcomes:
\begin{itemize}
    \item The predicted model discrepancy is very similar to the actual model discrepancy. This means that the error here in the calibration data is totally caused by model discrepancy, and tuning $\bm \theta$ does not help so this case is not informative for the calibration purpose.
    \item The predicted model discrepancy is not the same with the actual model discrepancy. This indicates that error can be caused by calibration parameters when model discrepancy is considered, thus the case can be informative to calibration.
\end{itemize}

\begin{figure}[!h]
    \centering
    \includegraphics[width = \textwidth]{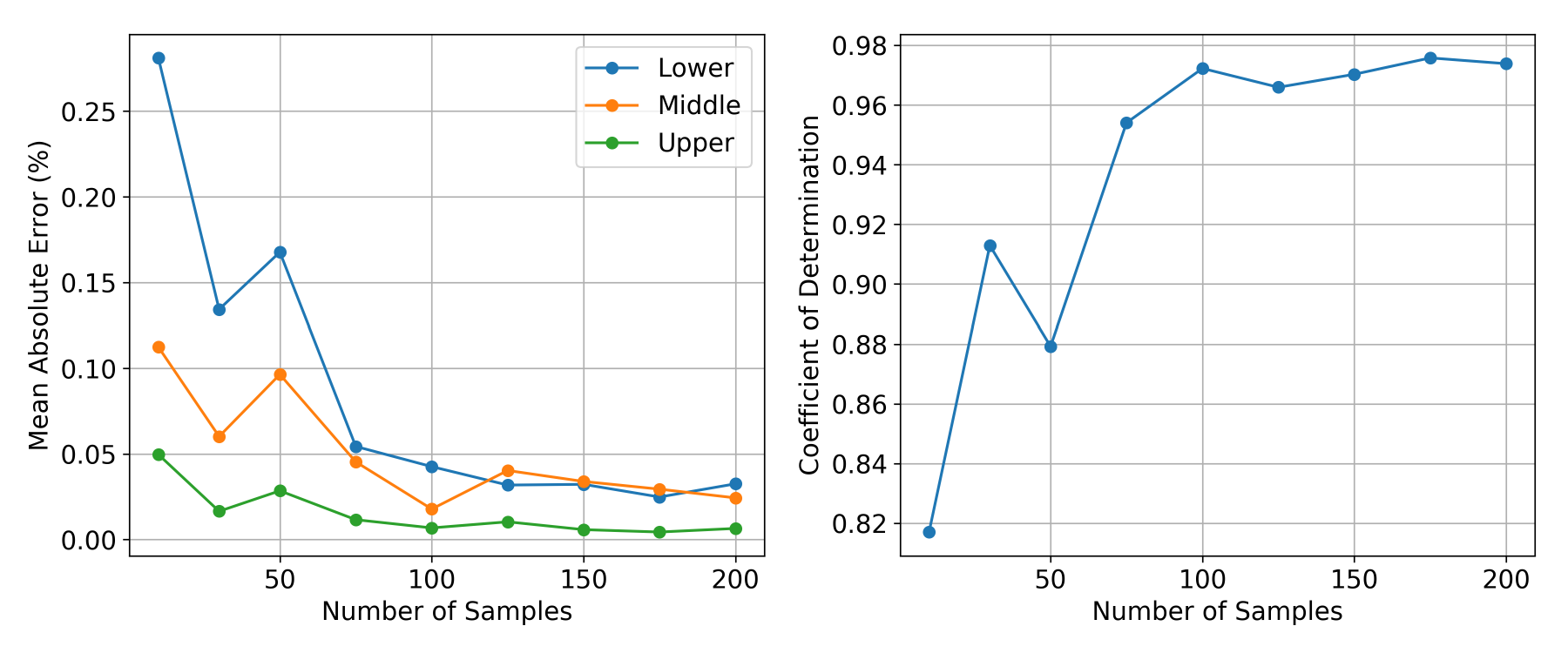}
    \caption{Convergence study for GP}
    \label{fig:val1}
\end{figure}

For the GP model of computer code $GP_{CC}$, extra sampling is required because its input includes $\bm \theta$ and we need to design how $\bm \theta$ is sampled to construct the GP model. For each $\bm_{cal}$ in the calibration set, we need a number of $\bm \theta$ samples from a certain range (0,5). The range can be adjusted larger if it is not sufficient for posterior distribution. The number of samples needs to be determined by the convergence study shown in Figure~\ref{fig:val1}. The mean absolute error of testing a randomly drawn testing dataset is quantified for GP models with an increasing number of samples. The samples are all drawn by the LHS method. We can see that 100 sample is typically sufficient to reach the lowest level of error and the highest level of coefficient of determination. So we use 100 samples for each case in the calibration set, and use vector ($\bm x_{cal}, \bm \theta$) as training date for $GP_{CC}$ to ensure its accuracy.

%% file: 5sec.tex
\section{Results of Modular Bayesian Model}
\label{sec5}

\subsection{Posterior Distribution}
Now that all elements in the posterior distribution of $\bm \theta$( Equation \ref{eqa:41}) have been collected, we can sample this posterior by MCMC. In this section, the adaptive MH algorithm with global scaling (algorithm \ref{algo:1}) is used. 20,000 samples are collected, and the first 4,000 are used as the burn-in period. Multiple chains are run and trace plots are examined to make sure the convergence of the chain.

\begin{equation}
\label{eqa:41}
    p(\bm \theta| \bm y^M, \mathcal{GP}_{CC}, \mathcal{GP}_{MD}) \propto p(\bm y^M| \bm \theta, \mathcal{GP}_{CC}, \mathcal{GP}_{MD} ) \cdot p(\bm \theta)
\end{equation}

\begin{figure}[!h]
    \centering
    \includegraphics[width = \textwidth]{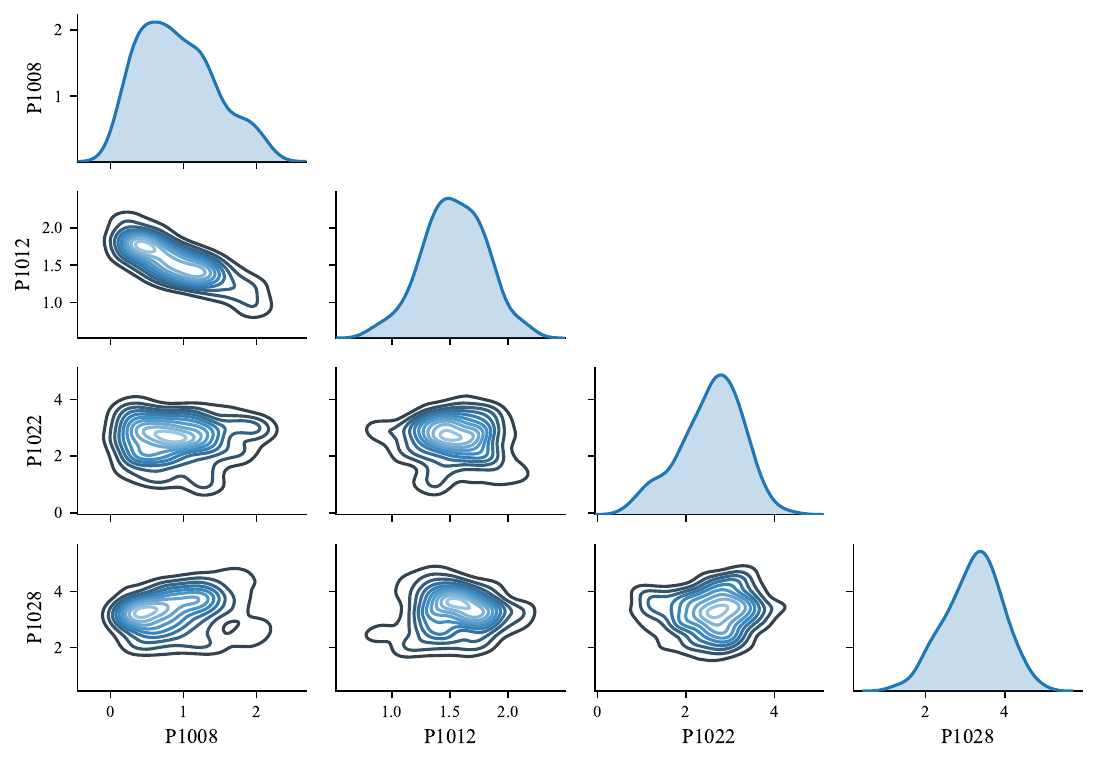}
    \caption{Posterior pair-wise joint and marginal distributions when model discrepancy is considered}
    \label{fig:post_bias}
\end{figure}

\begin{figure}[!h]
    \centering
    \includegraphics[width = \textwidth]{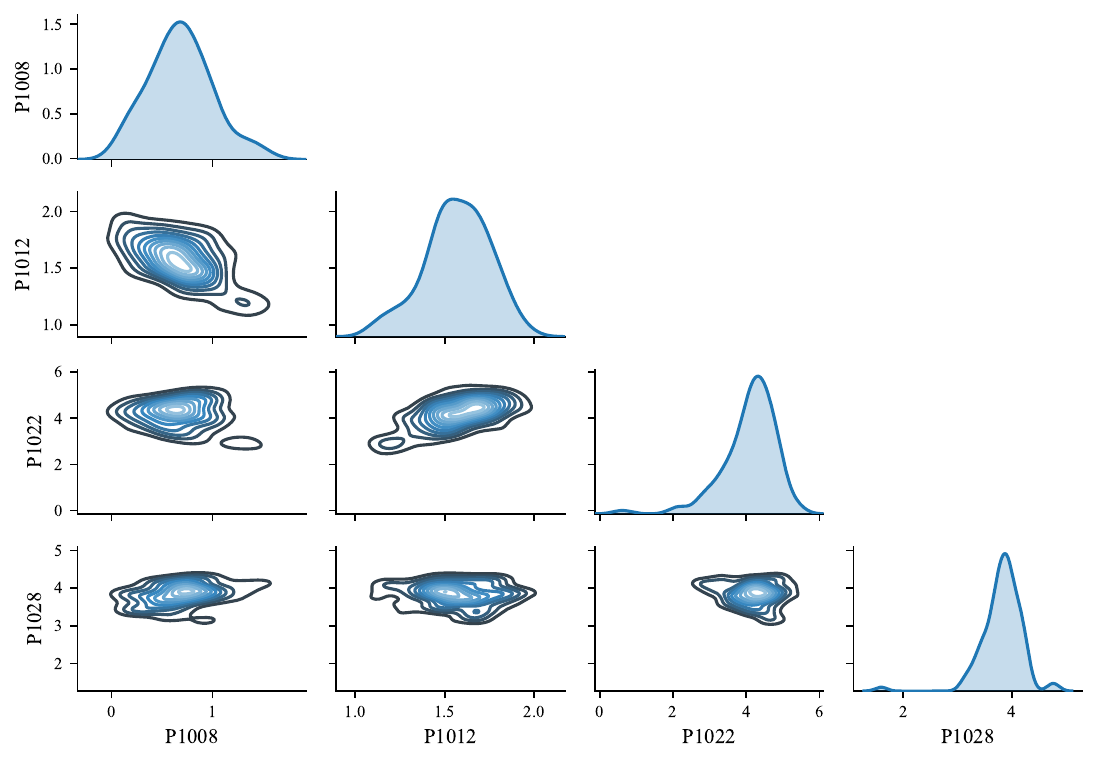}
    \caption{Posterior pair-wise joint and marginal distributions when model discrepancy is NOT considered}
    \label{fig:post_bias_no}
\end{figure}

\begin{figure}[!h]
    \centering
    \includegraphics[width = \textwidth]{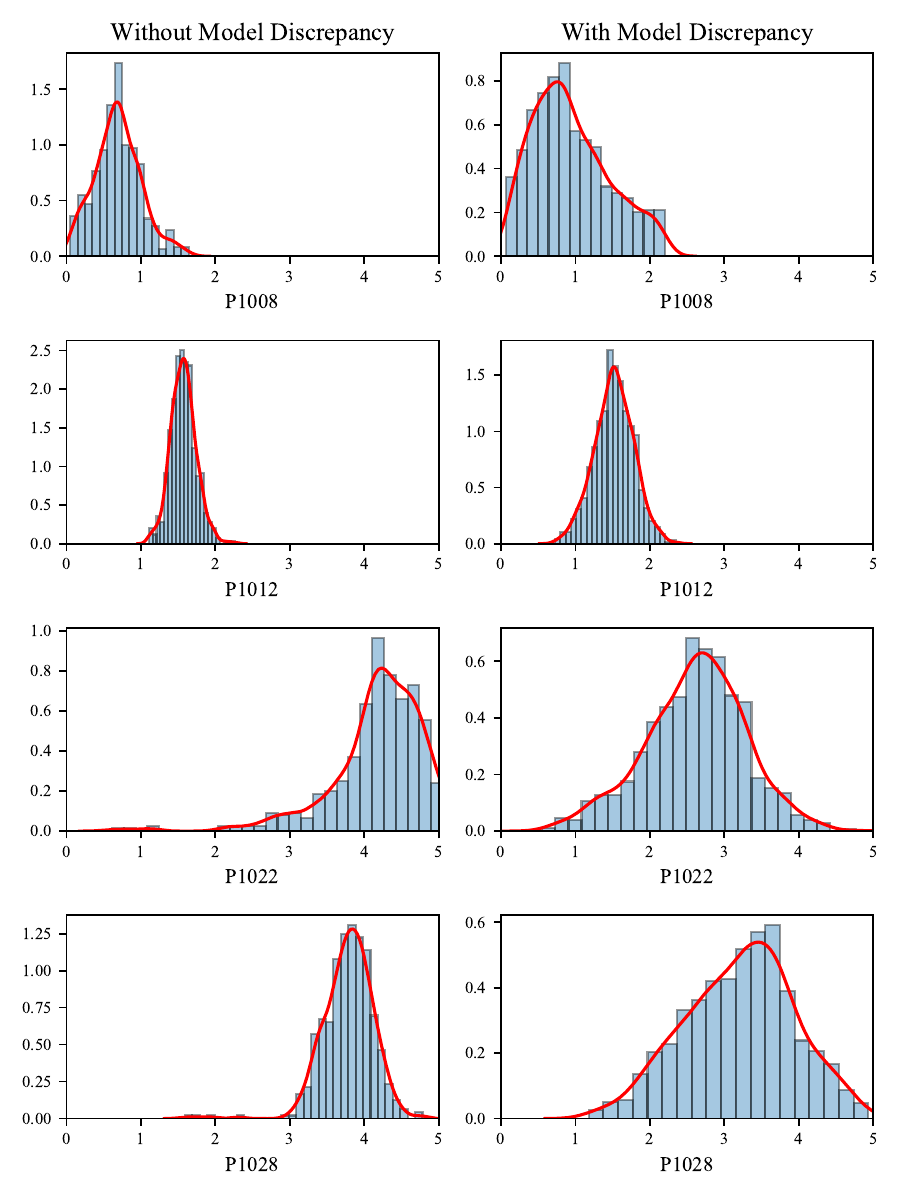}
    \caption{Posterior distributions by Modular Bayesian approach with and without model discrepancy}
    \label{fig:post_compare_bais}
\end{figure}

At the same time, for comparison purposes, the calibration was also conducted when the model discrepancy is not considered. In this case, the same calibration set $\bm x_{cali}$ (20 cases) is used to ensure the experimental information is the same, and there is no need to model $GP_{MD}$ term, which makes things easier. The same MCMC algorithm is applied to the case with no model discrepancy term, and posterior samples are obtained. Figure \ref{fig:post_bias} and \ref{fig:post_bias_no} show the posterior pair-wise joint and marginal distributions when model discrepancy is and is not considered, respectively. We can see that both results show an obvious correlation between parameter `P1008' and `P1012', and all posterior show normal or normal-like shapes. The negative correlation coefficient between these two parameters is also consistent with the physical phenomenon, as both will lead to higher void fraction value when they get larger. We should note that the x-axes are different in two figures, the posteriors when the model discrepancy is not considered are narrower. This fact can be observed in Figure \ref{fig:post_compare_bais}, which shows the marginalized posterior for each calibration parameter.

The plots in the left column of Figure \ref{fig:post_compare_bais} are posteriors calibrated without model discrepancy, and the plots in the right column are calibrated with the model discrepancy. 

As mentioned before, ignoring the presence of model discrepancy in Bayesian calibration will typically lead to over-fitting of the calibration parameters, and may make calibration parameters compensate for the model discrepancy unrealistically. This phenomenon can be seen from the comparison in Figure 4. The posterior standard deviations are smaller and pair-wise joint distributions are more concentrated when model discrepancy is not considered. Although this phenomenon might be preferable in some cases because it reduces prior uncertainty, it is also an indication of potential over-fitting because the concentration may be caused by the fact that a parameter is compensating for the model discrepancy. The figures considering model discrepancy can be seen as corrected by our available knowledge for the model discrepancy.

\subsection{Results Validation and Discussion}
\begin{figure}[!h]
    \centering
    \includegraphics[width = \textwidth]{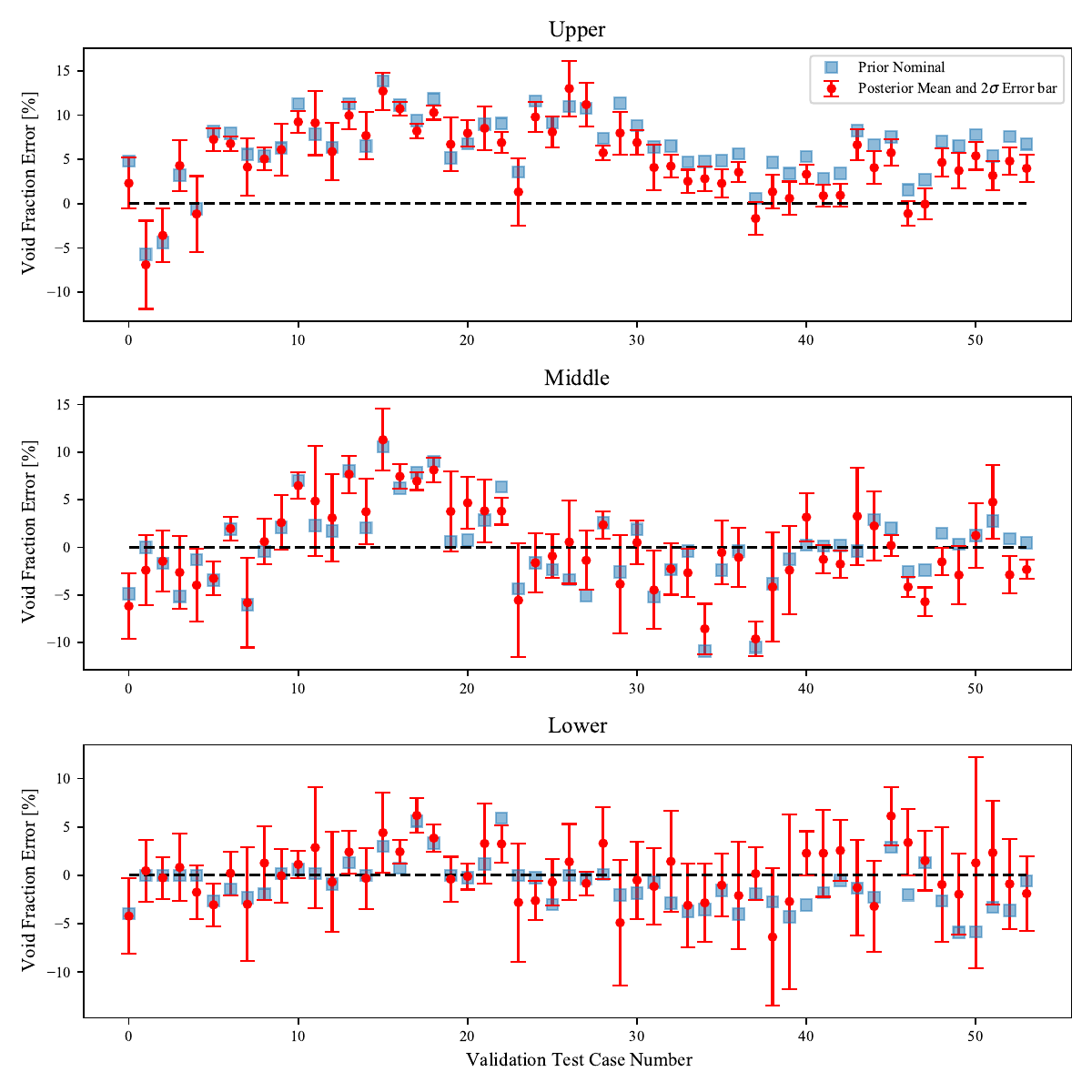}
    \caption{Comparison of TRACE posterior means (red dots) and TRACE output with prior nominal values (blue squares). The error bars represent the 95\% confidence interval of the model responses for each case}
    \label{fig:4vali}
\end{figure}

Since we do not know the underlying true value of these calibration parameters, the posteriors can be validated by checking if the simulation model with posterior $\bm \theta$ leads to better consistency with experiment data, on the validation dataset. Note that we have used 20 experiment cases for calibration so the rest cases 54 cases will be used for validation. The posterior distributions of model responses $\bm y^M(\bm x, \bm \theta_{post})$ given $p(\bm \theta \mid \bm y^E )$ can be calculated by integrating $\bm y^M(\bm x, \bm \theta_{post})$ with respect to $p(\bm \theta \mid \bm y^E )$. This can be done by Monte Carlo simulation using the posterior samples of $\bm \theta$. The mean and standard deviation of the model response posterior $\bm y^M(\bm x, \bm \theta_{post})$ is compared with the prior nominal of $\bm y^M(\bm x, \bm \theta_{prior})$ in Figure \ref{fig:4vali}. $\bm \theta_{prior}$ is taken as 1.0 here. The y-axis shows the void fraction error which is the difference between experimental and simulated void fraction. We can see that posterior means (red circles in Figure \ref{fig:4vali}) for these validation cases are generally closer to experimental data than the original prediction results, especially in the `Upper' measurement location.

The Root Mean Square Error (RMSE) between simulation results with experiment data for all validation cases are calculated in three conditions: (1) original predictions $\bm y^M(\theta = 1)$, (2) posterior of model response considering model discrepancy using modular Bayesian approach ($\bm y^M(\theta_{post})$), and (3) posterior of model response without model discrepancy. The results are reported in Table 

\begin{table}[h!]
    \centering
    \caption{Validation of calibration results by RMSE }
    \begin{tabular}[c]{l  c}
    \hline
    \textbf{Type of response} & \textbf{RMSE [\%]}\\
     \hline
     $\bm y^M(\theta = 1)$ & 5.19\\
     $\bm y^M(\theta_{post})$ with discrepancy & 4.73\\
     $\bm y^M(\theta_{post})$ no discrepancy & 5.23\\
    \hline
    \end{tabular}
    \label{tab:4vali}
\end{table}

We can see that when model discrepancy is not considered, the prediction accuracy is not improved in the validation set, indicating the previous results in Figure \ref{fig:post_bias_no} can be over-fitting.

%% file: 6sec.tex
\section{Summary}
\label{sec6}

Input uncertainty is an essential element in performing probabilistic uncertainty analysis in BEPU approaches.  A majority of the present BEPU methodologies are based on the propagation of uncertainties from inputs to outputs of predictive models, so the determination and justification of the uncertainty range associated with each uncertain parameters are necessary. The parametric uncertainty caused by empirical equations of state and constitutive equations (closure laws) in TH codes has been primarily addressed by ``expert judgment'' or ``user self-evaluation''. 

The variance-based Sobol method provides an intuitive tool for quantifying influential input parameters. The Sobol method based SA is conducted for TRACE physical model parameters in BFBT and PSBT benchmarks, respectively. 4 parameters that have impacts on QoIs are selected for calibration and will be treated as the input of the surrogate model.

Surrogate models are convenient tools to overcome the long time consumed by TH code runs. Various regression models can be used to construct the surrogate. Gaussian Processes have been widely used in the Bayesian calibration community due to its probabilistic nature and it capability of regressing complex input-output relationships using limited parameters (because GP is a non-parametric method). The modular Bayesian approach is applied to the steady-state PSBT data. The resulting input uncertainties are shown to be more consistent with available experimental data and can be used to replace the expert judgment in future forward UQ or SA analysis.